\documentclass{article}
\usepackage[cmex10]{amsmath}

\usepackage{authblk}

\usepackage{amsfonts}
\usepackage{amssymb}
\usepackage{tikz}
\usepackage{graphicx}
\usepackage{epstopdf}
\usepackage{mathtools}
\usepackage{float}
\usepackage{empheq}
\usepackage{amsthm}
\usepackage{caption}
\usepackage[font=normalsize]{subcaption}
\usepackage{enumitem}
\usepackage{algorithm}
\usepackage[noend]{algpseudocode}
\theoremstyle{definition}

\newtheorem{remark}{Remark}
\newtheorem{proposition}{Proposition}

\def\bbR{\mathbb{R}}

\def\lmax{\lambda_{\rm max}}

\def\zer{0_{n\times n}}
\def\nablat{\nabla_{\rm t}}
\def\nabsq{\nabla^2_{\rm t}}
\def\tmin{t_{\rm min}}
\def\tmax{t_{\rm max}}

\definecolor{bostonuniversityred}{rgb}{0.8, 0.0, 0.0}
\definecolor{cerulean}{rgb}{0.0, 0.48, 0.65}

\newcommand\BibTeX{{\rmfamily B\kern-.05em \textsc{i\kern-.025em b}\kern-.08em
T\kern-.1667em\lower.7ex\hbox{E}\kern-.125emX}}

\author[1,2]{Fairouz Zobiri}

\author[1]{Nacim Meslem}

\author[2]{Brigitte Bidegaray-Fesquet}

\affil[1]{Univ. Grenoble Alpes, CNRS, Grenoble INP, GIPSA-lab, 38000, Grenoble, France}

\affil[2]{Univ. Grenoble Alpes, CNRS, Grenoble INP, LJK, 38000, Grenoble, France}

\begin{document}

\title{Self-triggered stabilizing controllers for linear continuous-time systems}




\abstract[Abstract]{Self-triggered control is an improvement on event-triggered control methods. Unlike the latter, self-triggered control does not require monitoring the behavior of the system constantly. Instead, self-triggered algorithms predict the events at which the control law has to be updated before they happen, relying on system model and past information.\\
In this work, we present a self-triggered version of an event-triggered control method in which events are generated when a pseudo-Lyapunov function (PLF) associated with the system increases up to a certain limit. This approach has been shown to considerably decrease the communications between the controller and the plant, while maintaining system stability. To predict the intersections between the PLF and the upper limit, we use a simple and fast root-finding algorithm. The algorithm mixes the global convergence properties of the bisection and the fast convergence properties of the Newton-Raphson method. \\
Moreover, to ensure the convergence of the method, the initial iterate of the algorithm is found through a minimization algorithm. }



\maketitle


\section{Introduction}\label{Intro}
For a long time, the implementation of continuous-time control tasks on digital hardware has been tied to the so-called Shannon-Nyquist theorem. This condition requires the sampling frequency of the continuous control signal to be relatively high in order to avoid aliasing phenomena. This in turn requires the sensors, controller and actuators to communicate at high speed, tasks that can be straining on communication channels, energy sources and processing units. With the establishment of event-triggered control, researchers and engineers alike realized the possibility of taking samples at a lower pace, provided the samples are non-uniformly distributed over time. Less samples means less interactions between the different blocks of the system, less demand on the communication channels and computation resources.\\

\noindent Event-triggered control, however, only half-solves the problem. Event-triggered control works by updating the control law only when the controlled system violates predefined conditions on its states or output. This implies monitoring the state of the system continuously, thus inducing the high frequency exchanges that we were trying to avoid. Monitoring the event-triggering conditions might also require extra circuitry that is often difficult, if not impossible to build into existing plants.\\

\noindent One way to cancel the need for constant monitoring of the state is to predict in advance the time instants at which the conditions on system behavior are infringed. For this, we use the system's model to predict the evolution of its states. Control strategies in which the times of the control update are known beforehand are the topic of self-triggered control, a variant of event-triggered control. Self-triggered control is most often encountered in the framework of discrete-time systems \cite{Durand2012}, \cite{Velasco2015} \cite{Kishid2018}. In \cite{MazoSelfTriggered}, the event-triggering conditions are developed in continuous-time, whereas the next execution time is found by setting a time horizon that is divided in sub-intervals. An event is then determined by checking the event-triggering conditions in each sub-interval. Continuous-time systems have also been studied in \cite{Kobayachi2014}, where the problem is treated as an optimal control problem, with the next sampling instant as a decision variable. The result is a non-convex quadratic programming problem which is then approximated by a convex problem. In \cite{Wang2009a} and \cite{Wang2009} the authors suggest a self-triggered control method that preserves the $\mathcal{L}_2$ stability of the system in the presence of disturbances. Furthermore, self-triggered control schemes have often been coupled with model predictive control, as both use the model to project the behavior of the system up to some future time \cite{Kobayashi2012}, \cite{Henriksson2015}.\\

\noindent In this work, we design a self-triggered control algorithm for continuous-time linear time-invariant (LTI) systems. The algorithm predicts the times at which the system's behavior will infringe some predefined performance measures. We consider that the system is functioning properly when a pseudo-Lyapunov function (PLF) of its states is below a predefined upper bound. The control law is updated when the PLF reaches this upper bound. Predicting the events analytically is a difficult task, and thus, the self-triggered control algorithm computes an approximation of the event times via a minimization algorithm followed by a root-finding algorithm. The root-finding algorithm detects the intersections between the PLF and the upper limit, but needs to be properly initialized to converge to the right value. To do this, we take advantage of the shape of the PLF between two events; after the control is updated, the PLF decreases for some time, reaches a minimum and then increases again. This local minimum is easily computed via a minimization algorithm, and provides a good initial iterate for  the root-finding algorithm.\\

\noindent This paper is divided as follows. In Section~\ref{sec:Pb_def}, we present the problem that we are solving and establish the mathematical formalism necessary to expose our method. Section~\ref{sec:ST_algo} is divided into two parts. In the first part, we present the minimization algorithm and explain the motivation behind why we need this stage. In the second part, we give the details of the root-finding algorithm. Finally, in Section~\ref{sec:Simu}, we validate the method through a numerical example.

\section{Problem Formulation}\label{sec:Pb_def}
In this section, we first summarize the event-triggered control algorithm introduced in \cite{Zobiri2017} Then we introduce a self-triggered algorithm that predicts the events generated by this event-triggered algorithm.\\

\noindent Consider the following LTI system
\begin{eqnarray}
\begin{aligned}
\label{eq:sys_begin}
\dot{x}(t) &= Ax(t) + Bu(t),\\
x(t_0) &= x_0.
\end{aligned}
\end{eqnarray}
We want to stabilize System~\eqref{eq:sys_begin} with the following control sequence
\begin{eqnarray}
\begin{aligned}
\label{eq:u_tk}
u(t_k) &= -K x(t_k), \\
u(t) &= u(t_k),\hspace{1.5cm}\forall t\in [t_k,t_{k+1}),
\end{aligned} 
\end{eqnarray}
where $K$ is the feedback gain, selected such that the matrix $A-BK$ is Hurwitz. The time instants $t_k$ represent the instants at which the control law has to be updated to satisfy predefined stability or performance criteria. The objective of a self-triggered control implementation is to predict the time sequence $t_k$, $k = 0,~1,~2, ...$ at which the value of the control is updated.\\

\noindent The closed-loop form of System~\eqref{eq:sys_begin} can be written in an augmented form, with augmented state $\xi_k(t) = [x(t),~e_k(t)]^T \in \bbR^{2n}$ in $[t_k,t_{k+1})$, with $e_k(t) = x(t) - x(t_k)$ 
\begin{eqnarray}
\begin{aligned}
\label{eq:aug_sys}
\dot{\xi}_k(t) = \left[\begin{array}{cc}
A-BK & BK\\
A-BK & BK
\end{array}\right] \xi_k(t) \eqqcolon \Psi ~ \xi_k(t),
\end{aligned}
\end{eqnarray}
where $0_n$ is the vector of zeros in $\bbR^n$.
The system of equations~\eqref{eq:aug_sys} admits a unique solution on the interval $[t_k,t_{k+1})$ 
\begin{equation}
\label{eq:sol_st5}
\xi_k(t) = e^{\Psi (t-t_k)} \xi_k(t_k),
\end{equation}
where $\xi_k(t_k) = \left[\begin{array}{cc}
x(t_k) & 0_n^T
\end{array}\right]^T$.\\

\noindent We define $I_k(t)$ as the indicator function 
\begin{eqnarray}
\begin{aligned}
I_k(t) = \left\lbrace\begin{array}{cc}
1,&~t\in[t_k,t_{k+1}),\\
0,&~\text{otherwise}.
\end{array}\right.
\end{aligned}
\end{eqnarray}
Then, for all $t$, the state of the augmented system is given by
\begin{equation}
\xi(t) = \sum_k \xi_k(t) ~I_k(t),
\end{equation}
with initial state 
\begin{equation}
\xi(t_0) = \left[\begin{array}{cc}
x_0 & 0_n^T
\end{array}\right]^T \eqqcolon \xi_0,
\end{equation}

\noindent In what follows, we designate $\xi_k(t)$ as $\xi(t)$ when the two can be distinguished from the context.\\

\begin{remark}
When $t\in[t_k,t_{k+1})$, System~\eqref{eq:sys_begin} is written in closed-loop form as $\dot{x}(t) = A x(t) - BK x(t_k)$, with a solution $x(t) = (e^{A(t-t_k)} - A^{-1} (e^{A(t-t_k)} - I) BK)x(t_k)$, which requires $A$ to be non-singular. For this reason, we have chosen to work with the augmented system~\eqref{eq:aug_sys}, which admits a solution for all $A$ and does not exclude any class of systems. The proposed approached is then applicable to all stabilizable systems.
\end{remark}
\noindent To determine the control sequence, we first need to define the performance criteria that we impose on the system. For this, we associate to the system a positive definite, energy-like function of the state, that we refer to as a pseudo-Lyapunov function or PLF and which takes the following form 
\begin{equation}
\label{eq:Lyap_st}
V(\xi(t)) = \xi(t)^T \left[\begin{array}{cc}
P & \zer\\
\zer & \zer
\end{array}\right] \xi(t) \equiv \xi(t)^T\ \mathcal{P}\ \xi(t),
\end{equation}
where $\zer$ is the $n\times n$ matrix of zeros, and $P$ is a positive definite matrix that satisfies the following inequality 
\begin{equation}
\label{eq:gevp_ineq}
(A-BK)^T P + P (A-BK) \leq -\lambda P,
\end{equation}
where $\lambda > 0$.\\

\noindent For the control sequence given by Equation~\eqref{eq:u_tk} to stabilize the system, the PLF associated with the system has to decrease along the trajectories of the system. In this work, however, we relax this condition and only require from the PLF to remain upper bounded by a user-defined strictly decreasing threshold. Let the function $W(t)$ be such an upper bound, then, the PLF has to satisfy
\begin{equation}
\label{eq:cond_on_plf}
V(\xi(t)) \leq W(t).
\end{equation}

\begin{figure}
\centering
\begin{tikzpicture}[xscale = 1.2,yscale = .7]
\draw[->] (0,3) -- (7,3);
\draw[->] (0,3) -- (0,10);
\draw (7,2.5) node {time};
\draw [color=blue, domain=0:6.8, dashed, thick] plot(\x,{9*exp(-.15*\x)});
\draw (-0.6,9) node {\textcolor{blue}{\footnotesize{threshold}}}; 
\draw [color=bostonuniversityred,domain=0:2.7] plot(\x,{1*(\x-1.7)^2+5});
\draw (-0.6,7.7) node {\textcolor{bostonuniversityred}{\footnotesize{$V(x(t))$}}}; 
\draw[dotted] (2.7,3) -- (2.7,6);
\draw (2.9,2.5) node {$t_k$};
\draw [color=bostonuniversityred,domain=2.7:4.3] plot(\x,{2*(\x-3.7)^2+4});
\draw[dotted] (4.3,3) -- (4.3,4.7);
\draw (4.5,2.5) node {$t_{k+1}$};
\draw [color=bostonuniversityred,domain=4.3:5.5] plot(\x,{3*(\x-5)^2+3.2});
\draw [color=bostonuniversityred,domain=5.5:6.2] plot(\x,{(\x-12)*(0.0889*\x-1.0888)}); 
\end{tikzpicture}
\caption{The pseudo-Lyapunov function and the upper limit.}
\label{fig:general_behavior}
\end{figure}
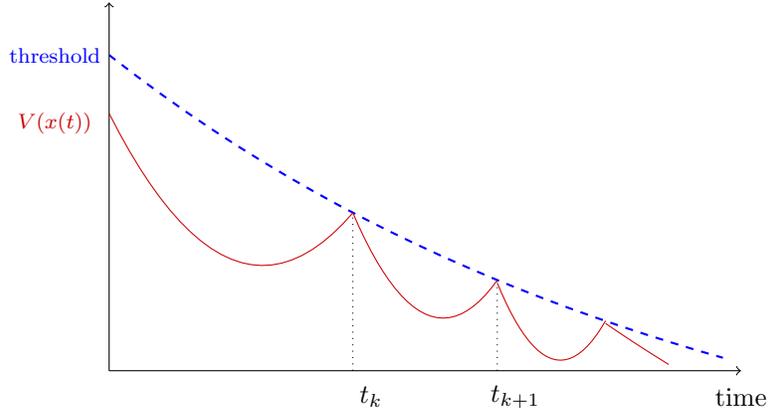

\noindent The upper bound $W(t)$ has to satisfy a few conditions. It has to be positive, strictly decreasing in time, and to ultimately tend toward zero. One suitable candidate is the exponentially decaying function of the form 
\begin{equation}
\label{eq:W_declare}
W(t) = W_0 e^{-\alpha (t-t_0)},
\end{equation}
where $W_0 \geq V(\xi(t_0))$ and $\alpha > 0$. The behaviors of $V(\xi(t))$ and $W(t)$ are depicted in Figure~\ref{fig:general_behavior}.\\
\noindent Furthermore, since we want to drive the system trajectory to equilibrium as fast as possible, and since the evolution of $V(\xi(t))$ is determined by the evolution of $W(t)$, we want $W(t)$ to decay to zero as fast as possible as well. The fastest possible rate of change of $W(t)$ is the largest scalar $\lambda$, that can be achieved from Inequality~\eqref{eq:gevp_ineq}, as shown in \cite{Zobiri2017} The largest possible value of $\lambda$ is the solution of the following generalized eigenvalue problem, 
\begin{eqnarray}
\begin{aligned}
\label{eq:gevp_prob}
&\text{maximize}& &\lambda\\
&\text{subject to}&  &(A-BK)^T P + P (A-BK) \leq -\lambda P, ~~P > 0.
\end{aligned}
\end{eqnarray}
Let $\lmax$ denote the solution of Problem~\eqref{eq:gevp_prob}. The rate of decay of $W(t)$ can be chosen as $0<\alpha<\lmax$.\\

\noindent Then, we can define the time instants $t_k$ as
\begin{equation}
\label{eq:events_def}
t_{k+1} = \inf \{t > t_k ~ |~ V(\xi(t)) = W(t)\}.
\end{equation}
with $t_0 = 0$.\\

\noindent In the next section, we detail the procedure used to predict the lower bounds of the entries of the time sequence $t_1,~t_2,~...$, knowing $t_0$.

\section{Self-triggered Algorithm}\label{sec:ST_algo}
Let $Z(t)$ denote the difference $W(t) - V(\xi(t))$. From Equation~\eqref{eq:events_def}, to determine $t_{k+1}$, it suffices to determine the successive time instants at which the following equation is verified
\begin{equation}
\label{eq:duh}
Z(t) = 0.
\end{equation}  
Equation~\eqref{eq:duh} depends on time and implicitly on the state $\xi(t)$ which depends on time through a transition matrix as seen from Equation~\eqref{eq:sol_st5}. This configuration renders Equation~\eqref{eq:duh} extremely difficult, if not impossible, to solve analytically. For this reason, we propose a numerical solution to Equation~\eqref{eq:duh}, where the instant $t_{k+1}$ is computed through a root-finding algorithm.\\

\noindent A numerical scheme needs an initial value, and our first guess would be to initialize the root-finding algorithm at instant $t_k$ in order to predict the instant $t_{k+1}$. However, the instant $t_k$ is itself a root, and as a result, the algorithm fails to converge to $t_{k+1}$ and finds $t_k$ as a solution again. Therefore, we have to initialize our algorithm at a later time instant. Let $\rho_k$ denote the first time instant at which the PLF reaches a local minimum after the time $t_k$. The instant $\rho_k$ is a good candidate for an initial value, and in what follows, we further justify its use in the root-finding algorithm.\\

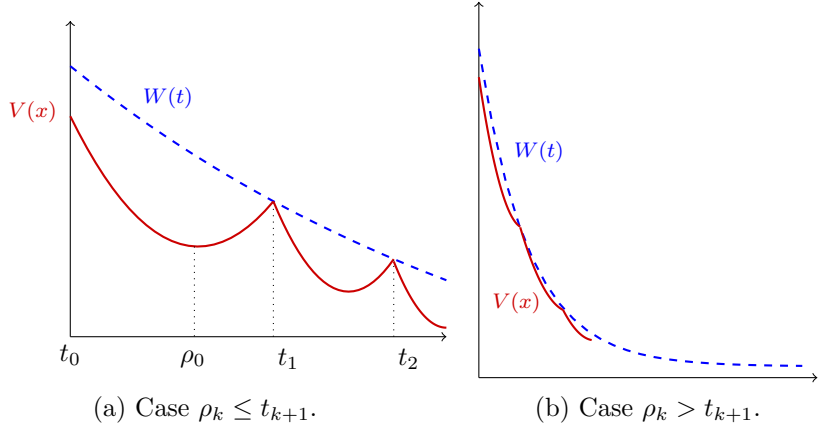
\begin{figure}[h!]
\centering
	\begin{subfigure}[t]{0.45\textwidth}
        \centering
        \begin{tikzpicture}[xscale = 1,yscale = .6]
\draw[->] (0,3) -- (5,3);
\draw[->] (0,3) -- (0,10);
\draw [color=blue, thick, domain=0:5, dashed] plot(\x,{9*exp(-.15*\x)});
\draw (1.3,8.3) node {\textcolor{blue}{\footnotesize{$W(t)$}}}; 
\draw [color=bostonuniversityred,thick,domain=0:2.7] plot(\x,{1*(\x-1.7)^2+5});
\draw (-0.5,8) node {\textcolor{bostonuniversityred}{\footnotesize{$V(x)$}}}; 
\draw [dotted] (1.65,5)--(1.65,2.894);
\draw (1.63,2.5) node {$\rho_0$};
\draw[dotted] (2.7,3) -- (2.7,6);
\draw (2.9,2.5) node {$t_1$};
\draw [color=bostonuniversityred,thick,domain=2.7:4.3] plot(\x,{2*(\x-3.7)^2+4});
\draw[dotted] (4.3,3) -- (4.3,4.7);
\draw (4.5,2.5) node {$t_2$};
\draw [color=bostonuniversityred,thick,domain=4.3:5] plot(\x,{3*(\x-5)^2+3.2});
\draw (0,2.6) node {$t_0$};
\end{tikzpicture}
		\caption{Case $\rho_k \leq t_{k+1}$.} \label{fig:case1}
    \end{subfigure} 
     \hspace{5mm}
	\begin{subfigure}[t]{0.4\textwidth}
        \centering
        \begin{tikzpicture}[xscale = 1,yscale = 1.]
\draw[->] (0.5,3) -- (5,3);
\draw[->] (0.5,3) -- (0.5,8);
\draw [color=blue, thick, domain=0.5:4.8, dashed] plot(\x,{9*exp(-1.512*\x)+3.15});
\draw [color=bostonuniversityred,thick,domain=0.5:1.05] plot(\x,{5.55*\x^2-12.22*\x+11.72});
\draw[color=blue] (1.3,6) node {\footnotesize{$W(t)$}};
\draw [color=bostonuniversityred,thick,domain=1.05:1.62] plot(\x,{2.5283797729618*\x^2 - 8.68059855521165*\x + 11.3270897832818});
\draw [color=bostonuniversityred,thick,domain=1.62:2.0] plot(\x,{2.31106612685559*\x^2 - 9.41869095816466*\x + 13.0931174089069});
\draw [color=bostonuniversityred] (1.,4) node {\footnotesize{$V(x)$}};
\end{tikzpicture}
		\caption{Case $\rho_k > t_{k+1}$.} \label{fig:case2}
    \end{subfigure}	
    \caption{Shape of the PLF for different choices of $\alpha$.}
\end{figure}

\noindent To do this, we classify the evolution of the PLF between two triggering instants, $t_k$ and $t_{k+1}$, into two categories. The first case, shown in Figure~\ref{fig:case1}, the minimum of the PLF occurs in between two consecutive triggering instants so that $t_k < \rho_k < t_{k+1}$. In this case, we can see that it is better to initialize our algorithm at time $\rho_k$, which when combined with the global properties of the bisection method, avoids a convergence toward the time $t_k$. In the second case, the PLF intersects with the threshold before reaching a local minimum (see Figure~\ref{fig:case2}). In this case the instant $\rho_k$ offers an upper bound on $t_{k+1}$ from which we can work our way backwards to recover the instant $t_{k+1}$.\\

\noindent Therefore, we need to precede the root-finding algorithm by a minimization stage, aimed at identifying the time instants at which $V(\xi(t))$ reaches a local minimum.

\subsection{Minimization Stage} \label{subsec:min_alg}
Once again, the complexity of the problem makes it impossible to synthesize a closed form analytical solution, and we suggest a numerical solution instead. The minimization algorithm is a modified Newton algorithm that uses $t_k$ as an initial guess to locate the minimum of $V(\xi(t))$ for $t>t_k$.\\

\noindent At each iteration, we compute the Newton step denoted by $\Delta \rho$. Let $\nablat V$ and $\nabsq V$ denote the first and second time derivatives of $V(\xi(t))$. Then, the Newton step is computed as
\begin{equation}
\label{eq:Newton_step}
\Delta \rho = \frac{-\nablat V}{|\nabsq V|}.
\end{equation} 
The expressions of $\nablat V$ and $\nabsq V$ are given by
\begin{align}
\nablat V &= \xi(t)^T \left[\begin{array}{cc}
M & L \\
L^T & \zer
\end{array}\right] \xi(t), \label{eq:nablaV}\\
\nabsq V &= \xi(t)^T \left[\begin{array}{cc}
\Lambda & \Gamma \\
\Gamma^T & \gamma
\end{array}\right] \xi(t), \label{eq:nablaSQV}
\end{align}
where $\xi(t)$ is given by Equation~\eqref{eq:sol_st5} and
\begin{flalign}
M &= (A-BK)^T P + P (A-BK), \nonumber\\
L &= PBK, \nonumber \\
\Lambda &= (A-BK)^T M + M(A-BK) + (A-BK)^T L^T+L(A-BK), \nonumber\\
\Gamma &= (A-BK)^T L + MBK + LBK, \nonumber\\
\gamma &= L^T B K + K^T B^T L. \nonumber
\end{flalign}

\noindent The minimization procedure is given in Algorithm~\ref{alg:min}. The current iterate is denoted as $\rho$ while the Newton step is represented by $\Delta \rho$. The number of iterations is bounded by the parameter $MaxIter$ for safety, in case the algorithm fails to converge. The procedure starts by computing a Newton step as given by Equation~\eqref{eq:Newton_step}. Then, a line search is performed to scale the Newton step. The Newton step is scaled such that the function $V$ decreases enough in the search direction. This step is needed because Newton's method for minimization is an algorithm that computes the roots of the first derivative of the function to be minimized. In our case, we have observed that the first derivative may contain an extremum near the root. Taking the tangent of $\nablat V$ at these points yields unreasonable Newton steps \cite{Gill1986} that need to be damped. For this reason, this method is sometimes referred to as the damped Newton's method \cite{Boyd2014}. Once the scaling factor is found, the damped Newton step is taken and a new iterate is found.\\

\begin{algorithm}[h!]
\caption{Minimization Algorithm} \label{alg:min}
\begin{algorithmic}[1]
\Function{Minimization($t_k$)}{}
\State $\rho$ $\gets$ $t_k$
\While {$\textit{iter} \leq \textit{MaxIter}$}
\State $\Delta \rho$ $\gets$ $-{\nablat V}/{|\nabsq V|}$
\State $s$ $\gets$ $1$
\While {$V(\xi(\rho+s\Delta \rho)) - V(\xi(\rho)) \geq  \kappa_1 \ \nablat V \ s \ \Delta \rho$}
\State $s \gets \beta s$, $\ \ \beta\in(0,1)$, $\kappa_1\in(0,0.5)$
\EndWhile
\State $\textit{tmp} \gets \rho$
\State $\rho$ $\gets$ $\rho + s \Delta \rho$
\If {$|\textit{tmp} - \rho|<tol$}
\State return $\rho_k = \rho$
\EndIf
\State $iter ++$
\EndWhile 
\State return $\rho_k$
\EndFunction
\end{algorithmic}
\end{algorithm}

\noindent Lines $5$ through $8$ of Algorithm~\ref{alg:min} correspond to a backtracking line search. The line search works as follows: a Taylor series approximation of $V(\xi(t))$ is computed, then the line search variable is decreased until a suitable reduction in $V(\xi(t))$ is achieved. The parameter $\kappa_1$ indicates the percentage by which $V(\xi(t))$ has to decrease along the search direction. The final value of $s$ is the quantity by which the Newton step is scaled, and $\beta$ is the fraction by which $s$ is decreased in each line search iteration.\\

\noindent Algorithm~\ref{alg:min} terminates when the change in $\rho$ from one iteration to the next becomes negligible. The algorithm's convergence can be very fast, first, because many time consuming operations can be carried out offline. This is the case for matrices $M$, $L$, $\Gamma$, $\Lambda$ and $\gamma$. Even the introduction of a backtracking line search, which is usually a time consuming procedure, does not slow down the algorithm. This is due to the fact that the line search is only performed when we are far from the minimizer, but becomes unnecessary as we approach the minimal value. Therefore, we noticed through our experiments that the algorithm's execution time is negligible compared to the length of the interval $t_{k+1} - t_k$.

\begin{remark}
In the case of one-dimensional systems, the times at which the local minima of $V(\xi)$ occur can be found analytically. The analytical expression for finding $\rho_k$ and its derivation are given in the Appendix. \\
When tested on numerical examples, the analytical expression and the numerical approach return the same time sequence.
\end{remark}

\subsection{Root-finding Algorithm}\label{subsec:root_alg}
Since we want our root-finding algorithm to be both fast and precise, we select an algorithm that combines Newton's method and the bisection method. The bisection method is a globally convergent method that acts as a safeguard against failures of the algorithm when we are far from the root. Newton's algorithm, on the other hand, has a quadratic convergence rate near the root and is used to speed up the algorithm. \\

\noindent To be able to use the bisection, we need to locate the root within an interval, that we denote $[\tmin,\tmax]$. This is a simple enough task once we know the time instant $\rho_k$. As explained earlier, $t_{k+1}$ can occur either before or after the time instant $\rho_k$. Either case is identified by computing $Z(\rho_k)$; if $Z(\rho_k)>0$, then $t_{k+1} > \rho_k$, whereas if $Z(\rho_k)<0$, $t_{k+1} < \rho_k$. We then define two time instants $t_1$ and $t_2$, we set $t_1 = \rho_k$  and  we follow the appropriate procedure
\begin{itemize}
\item Case $t_{k+1} > \rho_k$:\\
We pick a parameter $\theta > 0$. We suggest to scale the value of $\theta$ on the time lapse $\rho_k - t_k$. The scaling factor $\kappa_2$ is chosen between $0$ and $0.5$, depending on how crude we want the search to be, resulting in $\theta = \kappa_2 (\rho_k - t_k)$. Then, starting from $t_2 = t_1 + \theta$, we keep increasing $t_2$ by a value $\theta$ until $Z(t_2) < 0$. This procedure is depicted in Figure~\ref{fig:prepro1}. Finally, we find $\tmin = t_1$ and $\tmax = t_2$.
\item Case $t_{k+1} < \rho_k$:\\
In this case, we pick $\theta = -\kappa_2 (\rho_k - t_k)$. Starting from $t_2 = t_1 + \theta$, and while $Z(t_2) < 0$, $\theta$ is decreased by a factor of $2$ and $t_2$ is decreased by a value $\theta$. This procedure is depicted in Figure~\ref{fig:prepro2}. We keep dividing $\theta$ by $2$ to avoid the situation $t_2 < t_k$ when the search is too crude. Then, we set $\tmin = t_2$ and $\tmax = t_1$.
\end{itemize}
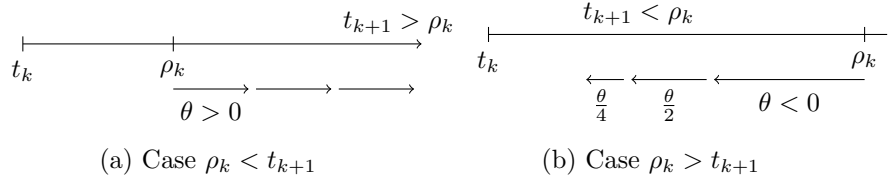
\begin{figure}[h!]
\centering
	\begin{subfigure}[t]{0.45\textwidth}
		\begin{tikzpicture}[xscale = 1,yscale = 1]
\draw[->] (1,0) -- (6.3,0);
\draw (6,0) node[above] {$t_{k+1} > \rho_k$};
\draw (1,-0.1) node[below] {$t_k$} -- (1,0.1);
\draw (3,-0.1) node[below] {$\rho_k$} -- (3,0.1);
\draw[->] (3,-0.6) -- (4,-0.6);
\draw (3.5,-0.9) node {$\theta>0$};
\draw[->] (4.1,-0.6) -- (5.1,-0.6);
\draw[->] (5.2,-0.6) -- (6.2,-0.6);
\end{tikzpicture}
		\subcaption{Case $\rho_k < t_{k+1}$}
		\label{fig:prepro1}
	\end{subfigure}
	\hspace{5mm}
	\begin{subfigure}[t]{0.4\textwidth}
		\begin{tikzpicture}[xscale = 1,yscale = 1]
\draw (1,0) -- (6.3,0) ;
\draw (3,0) node[above] {$t_{k+1} < \rho_k$};
\draw (1,-0.1) node[below] {$t_k$} -- (1,0.1);
\draw (6,-0.1) node[below] {$\rho_k$} -- (6,0.1);
\draw[->] (6,-0.6) -- (4,-0.6);
\draw (5,-0.9) node {$\theta<0$};
\draw[->] (3.9,-0.6) -- (2.9,-0.6);
\draw[->] (2.8,-0.6) -- (2.3,-0.6);
\draw (3.4,-0.6) node[below] {$\frac{\theta}{2}$};
\draw (2.5,-0.6) node[below] {$\frac{\theta}{4}$};
\end{tikzpicture}
		\subcaption{Case $\rho_k > t_{k+1}$}
		\label{fig:prepro2}
	\end{subfigure}
\caption{Locating the root inside an interval.}
\label{fig:prepro}
\end{figure}

\noindent The pre-processing stage is synthesized in Algorithm~\ref{alg:interval}.
\begin{algorithm}[h!]
\caption{Interval Finding} \label{alg:interval}
\begin{algorithmic}[1]
\Function{Pre-processing($t_k$)}{}
\State $t_1 \gets \rho_k$
\If {$Z(t_1) < 0$}
\State $\theta \gets -\kappa_2 (\rho_k - t_k), \ 0<\kappa_2\leq 0.5$
\Else 
\State $\theta \gets \kappa_2(\rho_k - t_k)$
\EndIf
\State $t_2 \gets \rho_k + \theta$
\While {$ Z(t_1) Z(t_2)\geq 0$}
\State $t_2 \gets t_2 + \theta$
\If {$t_2 \leq t_k$}
\State $t_2 \gets t_2 - \theta$, $\theta \gets \theta/2$
\State $t_2 \gets t_2 + \theta$
\EndIf
\EndWhile
\State $\tmax \gets \max(t_1,t_2)$
\State $\tmin \gets \tmax - |\theta|$
\State return $\tmin$, $\tmax$
\EndFunction
\end{algorithmic}
\end{algorithm}

\noindent The root-finding algorithm can only find approximate event times $t_k$, and so at $t = t_k$, we only have $W(t_k) \approx V(\xi(t_k))$. For this reason, to ensure the convergence of the algorithm, at $t=t_k$, we make the correction $W(t_k) = V(\xi(t_k))$. If we let $W(t_k) = W_k$, the expression of $W(t)$ on the interval $[t_k,t_{k+1})$ becomes
\begin{equation}
\label{eq:wk}
W(t) = W_k e^{-\alpha (t-t_k)},
\end{equation}
The function $Z(t)$, on $[t_k,t_{k+1})$, is then given by the 
\begin{equation}
\label{eq:Z(t)}
Z(t) = W_k e^{-\alpha(t - t_k)} - \xi(t)^T\ \mathcal{P}\ \xi(t),
\end{equation}
where $\xi(t)$ is given by equation~\eqref{eq:sol_st5}.\\
The first derivative with respect to time, along the trajectories of $\xi(t)$ is 
\begin{align}
\label{eq:dZdt}
\frac{dZ(t)}{dt} = -W_k \alpha & e^{-\alpha(t-t_k)} - \nablat V.
\end{align}

To decide whether to take a Newton step or a bisection step, we first compute an iterate with Newton's method. If the new iterate is located within the previously identified interval $[\tmin,\tmax]$, it is accepted. Otherwise, the Newton iterate is rejected and instead a bisection iterate (the mid-point of the search interval) is computed. The interval $[\tmin,\tmax]$ is then updated.

\noindent Algorithm~\ref{alg:root} describes the root-finding procedure in details.  It is a slightly modified version of the hybrid Newton-bisection algorithm found in \cite{Press2007}. To make the notations shorter, from now on we refer to $dZ(t)/dt$ as $\nablat Z(t)$.\\

\begin{algorithm}[h!]
\caption{Root-Finding Algorithm} \label{alg:root}
\begin{algorithmic}[1]
\Function{Newton-Bisection($\tmin,\tmax$)}{}
\If {$Z(\tmin) == 0$}
\State return $\tmin$
\EndIf
\If {$Z(\tmax) == 0$}
\State return $\tmax$
\EndIf
\State $t \gets (\tmin+\tmax)/2$
\State $\Delta t \gets \tmax-\tmin$,\hspace{5mm} $\Delta t_{\rm old} \gets \Delta t$
\State compute $Z(t)$, $\nablat Z(t)$
\While {$\textit{iter} \leq \textit{MaxIter}$}
\State $ step \gets \frac{Z(t)}{\nablat Z(t)}$
\If {$\tmin \geq t - step\ or \ \tmax \leq t - step\ or \ \frac{|\Delta t_{\rm old}|}{2} < |step|$}
\State $\Delta t_{\rm old} \gets \Delta t$
\State $\Delta t \gets (\tmax-\tmin)/2$
\State $t \gets \tmin + \Delta t$
\Else
\State $\Delta t_{\rm old} \gets \Delta t$
\State $\Delta t \gets step$
\State $t \gets t-\Delta t$
\EndIf
\If {$|\Delta t| < tol_2$}  \hspace{5mm} return $t$
\EndIf
\If {$Z(t) > 0$} \hspace{3mm} $\tmin \gets t$ \Else \hspace{3mm} $\tmax \gets t$
\EndIf
\EndWhile
\State return $t_{k+1}$
\EndFunction
\end{algorithmic}
\end{algorithm}
\noindent The algorithm starts by making sure that neither $\tmin$ nor $\tmax$ are the root, the procedure is exited if it is the case. Checking whether $\tmin$ is a root or not should be performed before the pre-processing, but for the sake of separation, we include it in the root-finding algorithm at this stage. The iterate $t$ is initialized as the midpoint of the interval $[\tmin,\tmax]$.\\

\noindent The variables $\Delta t$ and $\Delta t_{\rm old}$ store the current and the former step lengths, respectively. We compute $Z(t)$ and $\nablat Z(t)$ in order to compute the Newton step. The condition on line 13 of Algorithm~\ref{alg:root} decides whether a Newton step is taken or rejected. If by taking the Newton step we exceed $\tmax$ or regress below $\tmin$ or if Newton's algorithm is too slow, the Newton step is rejected, and a bisection step is taken instead. Lines 14 to 16 represent a bisection step, whereas lines 18 to 20 represent the case where the Newton step is taken.\\

\noindent After the new iterate is computed, we evaluate $Z(t)$ at that point. If $Z(t)$ is positive, the new iterate is located before the root, and it becomes $\tmin$. Otherwise, the current iterate become $\tmax$. The algorithm terminates when the change in $t$ between two consecutive iterates is too small, i.e. when the step length becomes smaller than a tolerance $tol_2$.
\subsection{Summary of the Self-Triggered Algorithm}
The three steps of the self-triggered algorithm, described separately so far, are grouped in the order in which they are called, in Algorithm~\ref{alg:all}. 
\begin{algorithm}
\caption{Self-Triggered Algorithm} \label{alg:all}
\begin{algorithmic}[1]
\Procedure{Self-triggered}{}
\State $\rho_k =$ MINIMIZATION ($t_k$)
\If {$Z(\rho_k) == 0$}
\State $t_{k+1} = \rho_k$
\EndIf
\State $[\tmin,\tmax]=$ PRE-PROCESSING ($\rho_k$)
\State $t_{k+1}=$ NEWTON-BISECTION($\tmin$, $\tmax$)
\EndProcedure
\end{algorithmic}
\end{algorithm}

\noindent The main contribution of this paper about the design of self-triggered stabilizing controllers is introduced in the following proposition
\begin{proposition}\label{prop:prop_only}
Let $\lmax$ be the solution to problem~\eqref{eq:gevp_prob}. If we choose $\alpha$ between $0$ and $|\lmax|$, Algorithm~\ref{alg:all} provides update instants $t_k$ for the control law $u(t)$, given by Equation~\eqref{eq:u_tk}, such that System~\eqref{eq:sys_begin} is asymptotically stable.
\end{proposition}
\noindent The proof for Proposition~\ref{prop:prop_only} is given in details in \cite{Zobiri2017}. In what follows, a brief summary of the proof is given. Since $W(t)$ decreases exponentially toward zero, we need to show that $V(\xi(t)) < W(t)$ for all $t$ (or equivalently that $Z(t) > 0$ for all $t$) to prove that System~\eqref{eq:sys_begin} is asymptotically stable. We know that in the interval $(t_{k-1},t_{k})$, $k \geq 1$, $Z(t) > 0$. And since Algorithm~\ref{alg:all} predicts the time $t_k$ when $Z(t)$ approaches zero from above, at $t=t_k$, the control law $u(t)$ is updated so that $Z(t)$ becomes strictly positive again. Therefore, $Z(t) > 0$ for all $t$.
\section{Numerical Simulation}\label{sec:Simu}
Consider the following third order LTI system \cite{Dorf2014},
\begin{equation*}
\dot{x}(t) = \left[\begin{array}{ccc}
1 & 1 & 0 \\
-2 & 0 & 4 \\
5 & 4 & -7 \\
\end{array}\right] x(t) + \left[\begin{array}{c}
-1 \\ 0 \\ 1
\end{array}\right]u(t),
\end{equation*}
with initial state $x_0 = [-2 \ 3 \ 5]^T$.\\
The system is unstable with poles at $-8.58$, $0.58$, $2.00$. We stabilize the system with a state-feedback control law with feedback gain
\begin{equation*}
K = \left[\begin{array}{ccc}
8.38  & 26.36  & 10.38
\end{array}\right],
\end{equation*}
that places the poles at $-1.14 \pm 1.35i$, $-5.71$. Solving the generalized eigenvalue problem~\eqref{eq:gevp_prob} yields $\lmax = 2.28$ and 
\begin{equation}
P = \left[\begin{array}{ccc}
275.7  & 1025.5  &  577.9\\
1025.5 &  3840.1  &  2173.5\\
577.9  & 2173.5   & 1234.1\\
\end{array}\right].
\end{equation}
We select $\alpha = 2.18~\rm s^{-1}$ and $W_0 = 1.3 V(x_0)$.\\
We simulate the system's operation for $7~\rm s$, with a sampling period $T_s = 10^{-3}$.\\

\noindent The values of the parameters required by the minimization algorithm and the root-finding algorithm are given in Table~\ref{tab:param_tab}.

\begin{table}[h!]
\caption{\label{tab:param_tab} Values of the parameters needed in the self-triggered control algorithm}
\centering
\begin{tabular}{lc}
\hline
Parameter & Value\\
\hline
$MaxIter$ & $50$\\
$\beta$ & $0.35$\\
$\kappa_1$ & $0.01$\\
$tol_1$ & $10^{-5}$\\
$\kappa_2$ & $0.25$\\
$tol_2$ & $10^{-5}$ at $t = 0$\\
\hline
\end{tabular}
\end{table}

\noindent The tolerance $tol_2$, at which the root-finding algorithm terminates, is set dynamically. Such a choice is motivated by the exponential decrease of $W(t)$, which tends to zero as time tends to infinity. If $tol_2$ is constant, at some point, $W(t)$ can decrease below this tolerance, and so does $V(\xi(t))$, leading to a small $Z(t)$ that could be mistaken for the root, when there is actually no intersection. Therefore, we index the value of $tol_2$ on $W_k$. As long as $W_k > 1$, $tol_2 = 10^{-5}$ as given, but W=when $W_k < 1$, then $tol_2$ is decreased according to the following equation
\begin{equation*}
tol_2 = 10^{-5-\phi}, \ \ \ \text{with}\ \  \phi = \lceil |\log_{10}(W_k)| \rceil.
\end{equation*}

\noindent At $t = 0$, we apply the control law $u(t_0) = -K x_0$ and we compute the instant $t_1$ using the self-triggered algorithm. The system is then on an open-loop configuration, only maintaining a control value of $u(t_0)$, until the clock signal displays the time $t_1$. At this point, the operation is repeated.\\

\begin{figure}[h!]
\centering
    \begin{subfigure}[t]{0.45\textwidth}
        \centering
        \includegraphics[height=50mm,width=\linewidth]{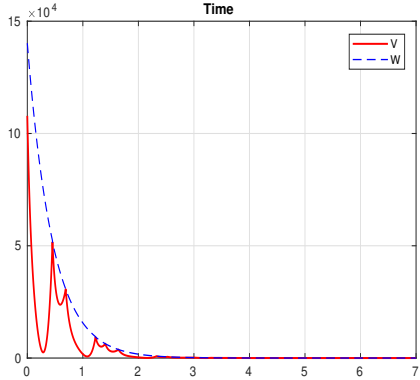} 
        \caption{PLF and threshold} \label{fig:lyap_st5}
    \end{subfigure}
    \hfill
    \begin{subfigure}[t]{0.45\textwidth}
        \centering
        \includegraphics[height=50mm,width=\linewidth]{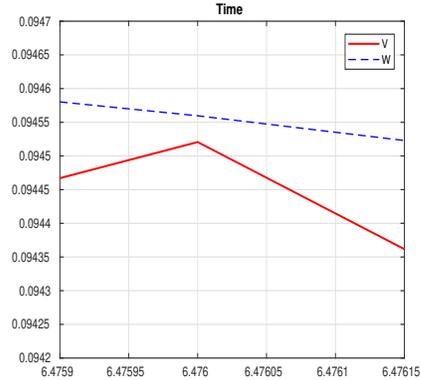} 
        \caption{Zoom on event at $t = 6.476~\rm s$} \label{fig:zoom_st5}
    \end{subfigure}

    \vspace{0.5cm}
    \begin{subfigure}[t]{0.45\textwidth}
    \centering
        \includegraphics[height=50mm,width=\linewidth]{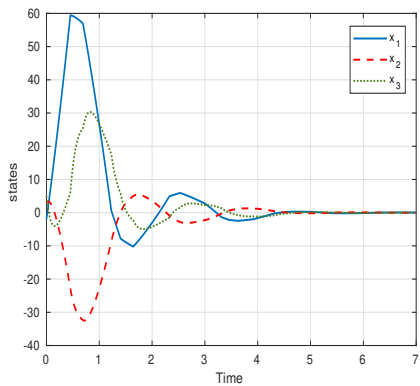} 
        \caption{States} \label{fig:states_st5}
    \end{subfigure}
    \hfill
    \begin{subfigure}[t]{0.45\textwidth}
    \centering
        \includegraphics[height=50mm,width=\linewidth]{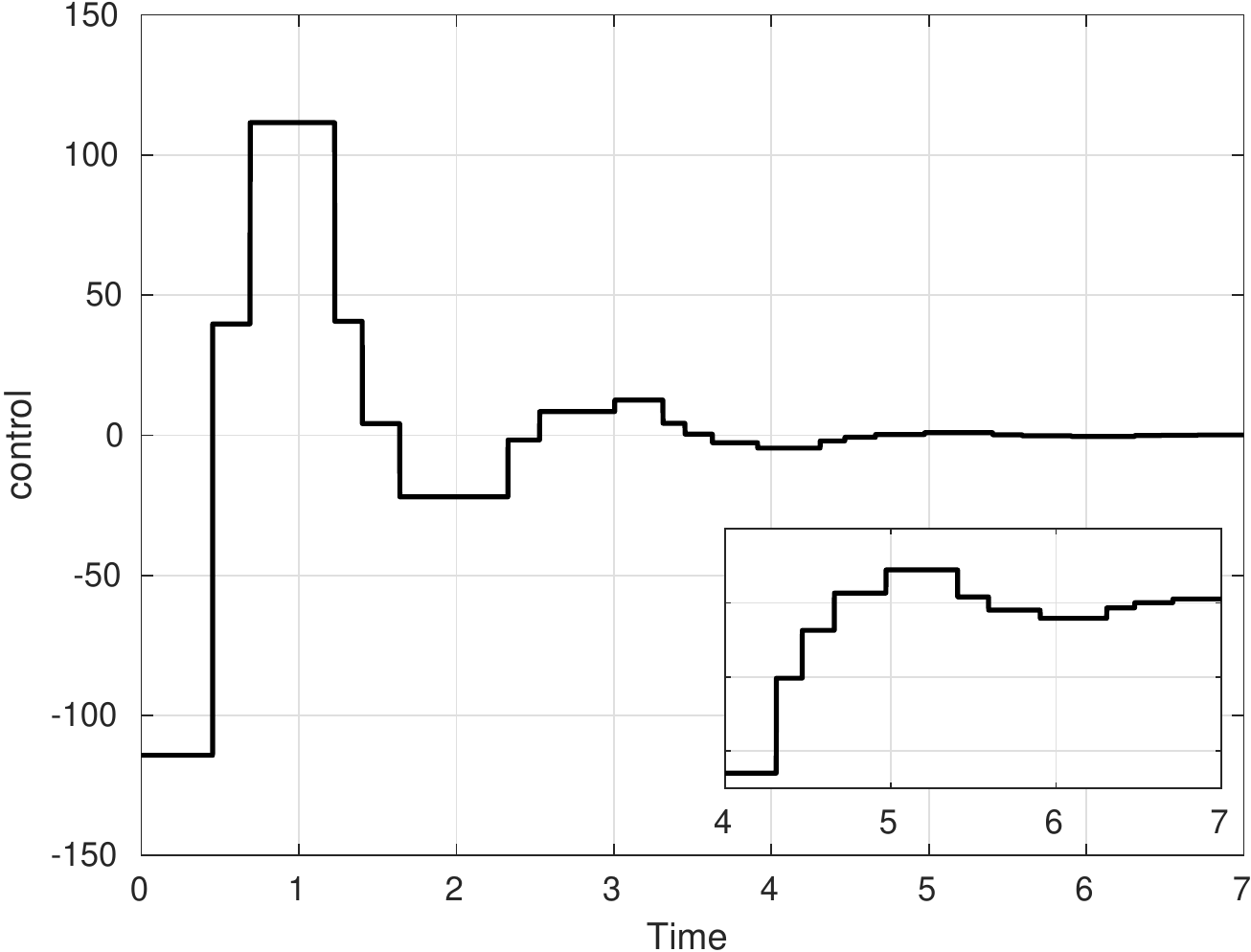} 
        \caption{Self-triggered control} \label{fig:control_st5}
    \end{subfigure}
    \caption{Simulation results of self-triggered control.}
\end{figure}

\noindent Figure~\ref{fig:lyap_st5} shows the time evolution of the functions $V(\xi(t))$ and $W(t)$. It shows that $V(\xi(t))$ remains below $W(t)$ at all times, which proves that the algorithm manages to identify correctly the times at which events occur, inducing an update of the control law. Even when the two functions approach zero, the intersections are still detected as shown on Figure~\ref{fig:zoom_st5}, which singles out an event at $t = 6.476~\rm s$ and $W(t) = 0.0948$.\\

\noindent The zoom on the event at $t = 6.476~\rm s$ shows that the update of the control law is carried out one time step before the intersection occurs. This is due to the fact that the control can only be updated at multiples of the simulation sampling period $T_s$. For this reason, when an intersection is predicted somewhere between sampling instants $t = 6.476~\rm s$ and $t = 6.477~\rm s$, we update the control law at the earlier instant, $t = 6.476~\rm s$, to prevent the PLF from crossing the threshold. By contrast, in the event-triggered control algorithm on which this approach is based, the event is detected one time step after it occurs. From this point of view, the self-triggered control algorithm represents another improvement on event-triggered control. \\

\noindent The three state variables, shown on Figure~\ref{fig:states_st5}, tend to equilibrium and the $\|x(t)\|$ stabilizes below $0.05$ within $ 6.94~\rm s$. The stabilizing control law is shown on Figure~\ref{fig:control_st5}. This figure shows the uneven distribution of updates in time. Figure~\ref{fig:control_st5} also includes a zoom on the control in the time interval $[4~\rm s,7~\rm s]$, which emphasizes the asynchronous nature of the updates, and which is not visible on the larger figure.\\

\begin{figure}
\centering
\includegraphics[width=110mm,height=50mm]{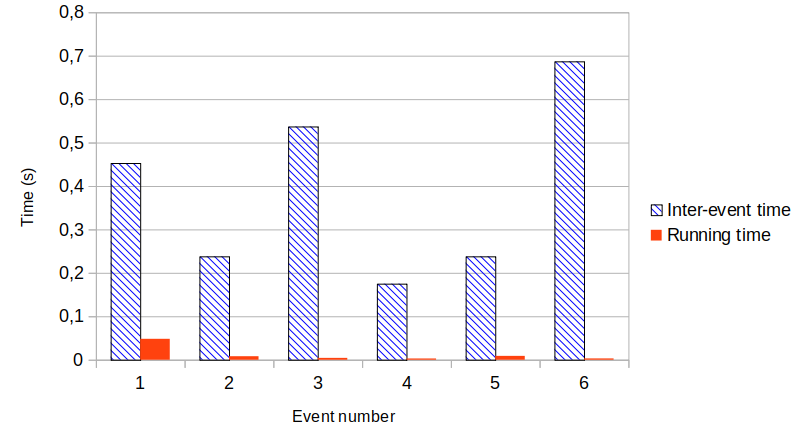}
\caption{The running times of the self-triggered algorithm versus the inter-event times.}
\label{fig:excel}
\end{figure}

\noindent Table~\ref{tab:run_times} lists the first six event times with the corresponding inter-event times $t_k-t_{k+1}$ and running times of the self-triggered control algorithm. We notice that for our experimental conditions, the algorithm's running time is much smaller than the corresponding inter-event time, allowing the online use of the algorithm. Moreover, the running time decreases as we go further in time, the highest running time being the first call of the algorithm, but this call can be made offline. Eventually, the running time settles around $0.002~\rm s$. Additionally, matrices $M$, $L$, $\Lambda$, $\Gamma$ and $\gamma$ are computed offline, and thus do not affect running time. Figure~\ref{fig:excel} further illustrates the disparity between the running times of the algorithm and the inter-event times.\\

\begin{table}[h]
\caption{The first $6$ events}
\centering
\begin{tabular}{ccc}
\hline
Update time & Inter-event time & Running time\\
\hline
0.453 & 0.453 & 0.0481\\
0.691 & 0.238 & 0.0081\\
1.228 & 0.537 & 0.0043\\
1.403 & 0.175 & 0.0029\\
1.641 & 0.238 & 0.0089\\
2.328 & 0.687 & 0.0030\\
\hline 
\end{tabular}
\label{tab:run_times}
\end{table}

\section{Conclusion}
We presented a self-triggered control algorithm for linear time-invariant systems. The approach approximately predicts the times at which a pseudo-Lyapunov function associated to the system reaches an upper limit, which are the times at which the system ceases to be stable and the control needs to be updated. These time instants are approximated using numerical methods in two stages. In the first stage, a minimization algorithm locates the time $\rho_k$ at which the pseudo-Lyapunov function reaches a minimum value in the interval between two events. In the second stage, a root-finding algorithm initialized at $\rho_k$ approximates the time of the next event.\\

\noindent The strength of this root-finding method is that it combines a globally convergent method with a locally convergent method. The globally convergent method ensures convergence to the right solution while the locally convergent method speeds up the convergence. Additionally, the minimization stage guarantees that the algorithm is initialized with a value close to the region of attraction of the actual root. The convergence and speed properties of this method make it suitable for both offline and online implementations. \\

\noindent To further validate this approach, the next step would be to apply the self-triggered control algorithm on a real system. This would allow us to test its efficiency against the uncertainties encountered in practical application. \\
Another perspective would be to extend this method to solve the problem of reference tracking, as this involves, in addition to stabilizing the system, the difficulty of detecting the changes in the reference trajectory. 

\section{Acknowledgment}
\noindent This work has been partially supported by the LabEx PERSYVAL-Lab (ANR-11-61 LABX-0025-01).

\appendix
\section{One-dimensional Systems}\label{app:one_d_sys}
In the case of one-dimensional systems, the local minimum of $V(\xi(t))$ can be found analytically. In what follows, we give a detailed procedure to determine $\rho_k$ analytically. We consider the first order LTI system described as
\begin{eqnarray}
\begin{aligned}
\label{eq:differential5}
\dot{x}(t) &= ax(t)+bu(t),\\
y(t) &=  c x(t),
\end{aligned}
\end{eqnarray} 
where $x(t),u(t) \in \bbR$, and $a, b, c \in {\bbR^*}$, $\forall t>0$.\\

Let $x_k$ denote $x(t_k)$. The event-triggered control law is given by $u(t) = -K x_k$ and System~\eqref{eq:differential5} in its closed-loop form is given by
\begin{equation}
\label{eq:cl_scalar}
\dot{x}(t) = a x(t) - bKx_k, \ \ \forall t\in [t_k,t_{k+1}),
\end{equation}

Since we assumed that $a\neq 0$, the augmented system described by Equation~\eqref{eq:aug_sys} is not needed for the scalar case. The differential equation~\eqref{eq:cl_scalar} admits a unique solution for $t>t_k$, given by 
\begin{equation}
\label{eq:state_sol_real}
x(t) = \left(\frac{bK}{a} +\left(1-\frac{bK}{a}\right) e^{a(t-t_k)}\right) x_k.
\end{equation}
To System~\eqref{eq:differential5}, we associate a Lyapunov-like function of the form
\begin{equation}
\label{eq:lyap_scalar}
V(x(t)) =  p x(t)^2,
\end{equation}
where $p > 0$ is a solution to the Lyapunov inequality
\begin{equation}
\label{eq:lyap_eqn_ch5}
2 p(a-bK) \leq -q,
\end{equation}
where $q > 0$ is a user-defined design parameter.\\

The minimum of $V(x(t))$ corresponds to
\begin{eqnarray}
\label{eq:lyap_zero_scal}
\begin{aligned}
0 = \frac{dV(x(t))}{dt} = 2p (a x(t) - b K x_k) x(t).
\end{aligned}
\end{eqnarray}

Equation \eqref{eq:lyap_zero_scal} admits two solutions, $x(t) = 0$, and $x(t) = bKx_k/a$. However, the solution $x(t) = bKx_k/a$ is impossible as it is equivalent to
\begin{eqnarray*}
\begin{aligned}
\left(\frac{bK}{a} +\left(1-\frac{bK}{a}\right) e^{a(t-t_k)}\right) x_k &= \frac{bKx_k}{a},\\
e^{a(t-t_k)} \left(1-\frac{bK}{a}\right) &= 0.
\end{aligned}
\end{eqnarray*}
We know that $e^{a(t-t_k)} \neq 0$ and we cannot choose $K$ such that $bK/a = 1$ or else we would destabilize the system.
Therefore, in the scalar case, $dV/dt = 0$, if and only if $x(t) = 0$.\\ 

Consequently, the local minima of $V(x(t))$ occur only when $x(t)=0$ and $\rho_k$ can be directly computed from Equation~\eqref{eq:state_sol_real} 
\begin{equation*} 
\label{eq:prelude_time}
\left(\left(1-\frac{bK}{a}\right) e^{a(\rho_k - t_k)} + \frac{bK}{a}\right)x_k = 0.
\end{equation*}
We know that $x_k \neq 0$, because at $t=t_k$, $V(x_k)=px_k^2 = W(t_k) \neq 0$, hence $x_k \neq 0$. Therefore, the times $\rho_k$ are given by the expression
\begin{equation}
\label{THE_time}
\rho_k = t_{k} + \frac{1}{a} \log\left(\frac{bK}{bK-a}\right).
\end{equation}
We can always take the logarithm of $bK/(bK-a)$ because this is always a positive quantity, as can be seen from the following proof.
\begin{itemize}
\item Case $a>0$ :\\
The feedback gain is chosen such that $a - bK < 0$. Then, $bK - a> 0$, and $bK > a > 0$. Since the numerator and denominator are both positive, then $bK/(bK-a) > 0$. Moreover, $bK/(bK-a) > 1$, proving that the $\rho_k$ computed by Equation~\eqref{THE_time} occurs indeed after $t_k$.
\item Case $a<0$ :\\
If the open-loop system is already stable, the objective of the control is certainly to place the pole further to the left. Then, the feedback gain is chosen such that $a-bK < a < 0$. Then, we must have $bK > 0$ and $bK-a > 0$. Consequently, as in the previous case, $bK/(bK-a) > 0$. Even if in this case $bK/(bK-a) < 1$, the $\rho_k$ given by Equation~\eqref{THE_time} still occurs after $t_k$.
\end{itemize}
Equation~\eqref{THE_time} is independent of $x_k$, indicating that the interval $[t_k,\rho_k]$ has the same length for all $k$.


\end{document}